\documentclass[
    ,final            
  ]
  {aipproc}

\layoutstyle{6x9}
\usepackage{amssymb}

\begin{document}

\title{Evidence From HETE-2 For GRB Evolution With Redshift}

\author{Carlo Graziani}{
  address={Department of Astronomy \& Astrophysics, University of Chicago},
}

\author{Donald Q. Lamb}{
  address={Department of Astronomy \& Astrophysics, University of Chicago}
}

\author{Takanori Sakamoto}{
  address={Department of Physics, Tokyo Institute of Technology}
  ,altaddress={RIKEN (Institute of Physical and Chemical Research)}
}

\author{Timothy Donaghy}{
  address={Department of Astronomy \& Astrophysics, University of Chicago}
}

\author{Jean-Luc Atteia}{
  address={Centre D'Etude Spatiale des Rayonnements, France}
}

\author{The HETE-2 Science Team}{
   address={
    An international collaboration of institutions including 
  MIT, LANL, U. Chicago, U.C. Berkeley, U.C. Santa Cruz (USA),
  CESR, CNES, Sup'Aero (France), RIKEN, NASDA (Japan), IASF/CNR (Italy),
  INPE (Brazil), TIFR (India) 
  }
}

\begin{abstract}

After taking into account threshold effects, we find that the
isotropic-equivalent energies $E_{\rm iso}$ and luminosities $L_{\rm
iso}$ of gamma-ray bursts (GRBs) are correlated with redshift at the 5\%
and 0.9\% signficance levels, respectively.  Our results are based on 10
{\it Beppo}SAX GRBs and 11 HETE-2 GRBs with known redshifts.  Our results
suggest that the isotropic-equivalent energies and luminosities of GRBs
increase with redshift.  They strengthen earlier clues to this effect
from analyses of the BATSE catalog of GRBs, using the variability of
burst time histories as an estimator of burst luminosities (and therefore
redshifts), and from an analysis of {\it Beppo}SAX bursts only.  If the
isotropic-equivalent energies and luminosities of GRBs really do 
increase with redshift, it suggests that GRB jets at high redshifts may
be narrower and thus the cores of GRB progenitor stars at high redshifts
may be rotating more rapidly.  It also suggests that GRBs at very high
redshifts may be more luminous -- and therefore easier to detect -- than
has been thought, which would make GRBs a more powerful probe of
cosmology and the early universe than has been thought.

\end{abstract}

\maketitle

\section{Introduction}

GRB sources are a cosmologically-distributed population.  Like all other
such populations, their properties presumably evolve with redshift.  In
this paper, we address two key questions:  what evidence exists bearing
on the evolution of GRB energetics as a function of redshift, and what is
the nature and magnitude of that evolution?

\section{Previous Indications of GRB Evolution}

\begin{figure}[t]

\hspace{-1.0in}
\begin{minipage}{2.7truein}
\includegraphics[width=2.65truein, clip]{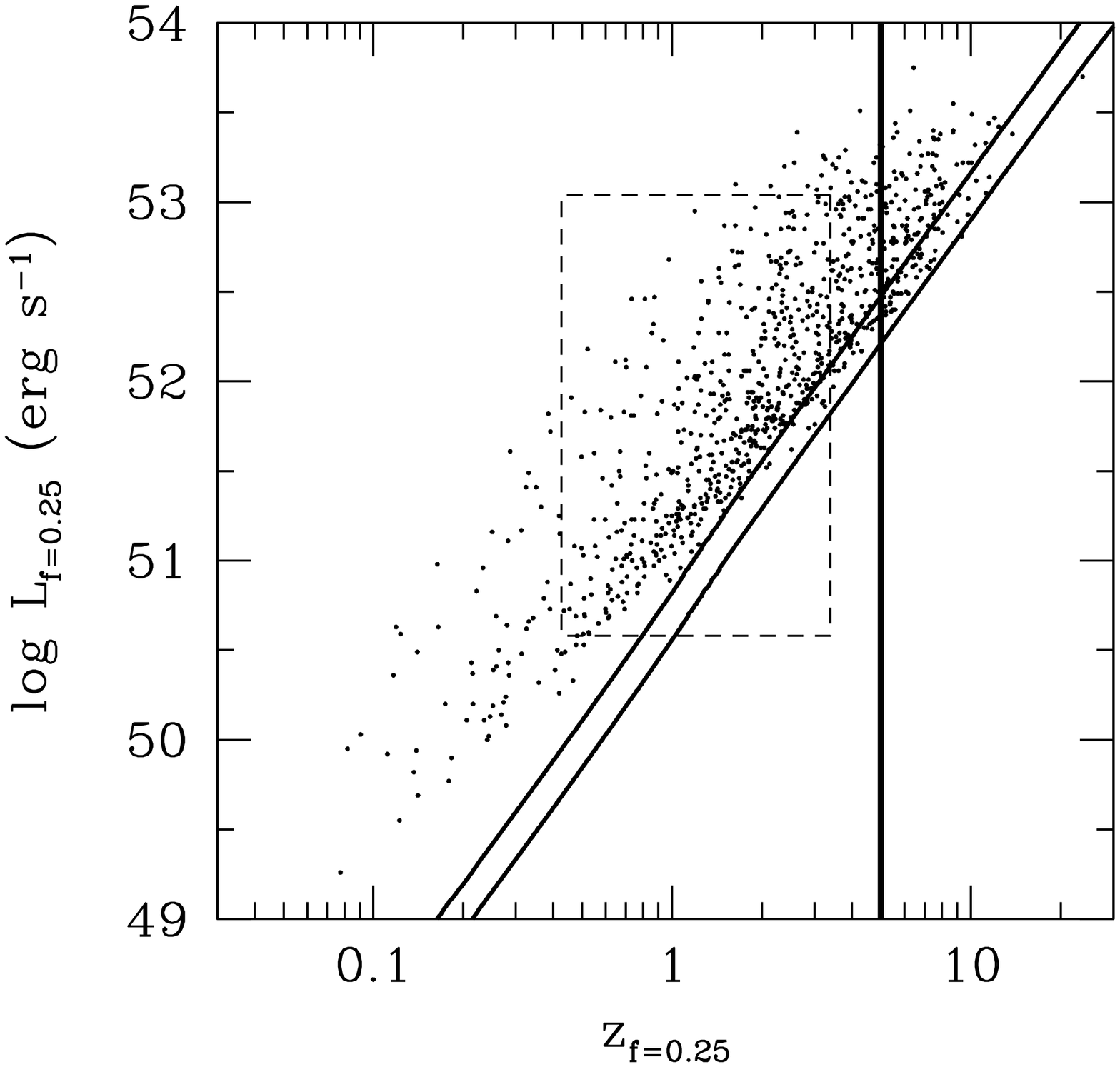}
\end{minipage}
\begin{minipage}{2.1truein}
\includegraphics[width=3.0truein, clip]{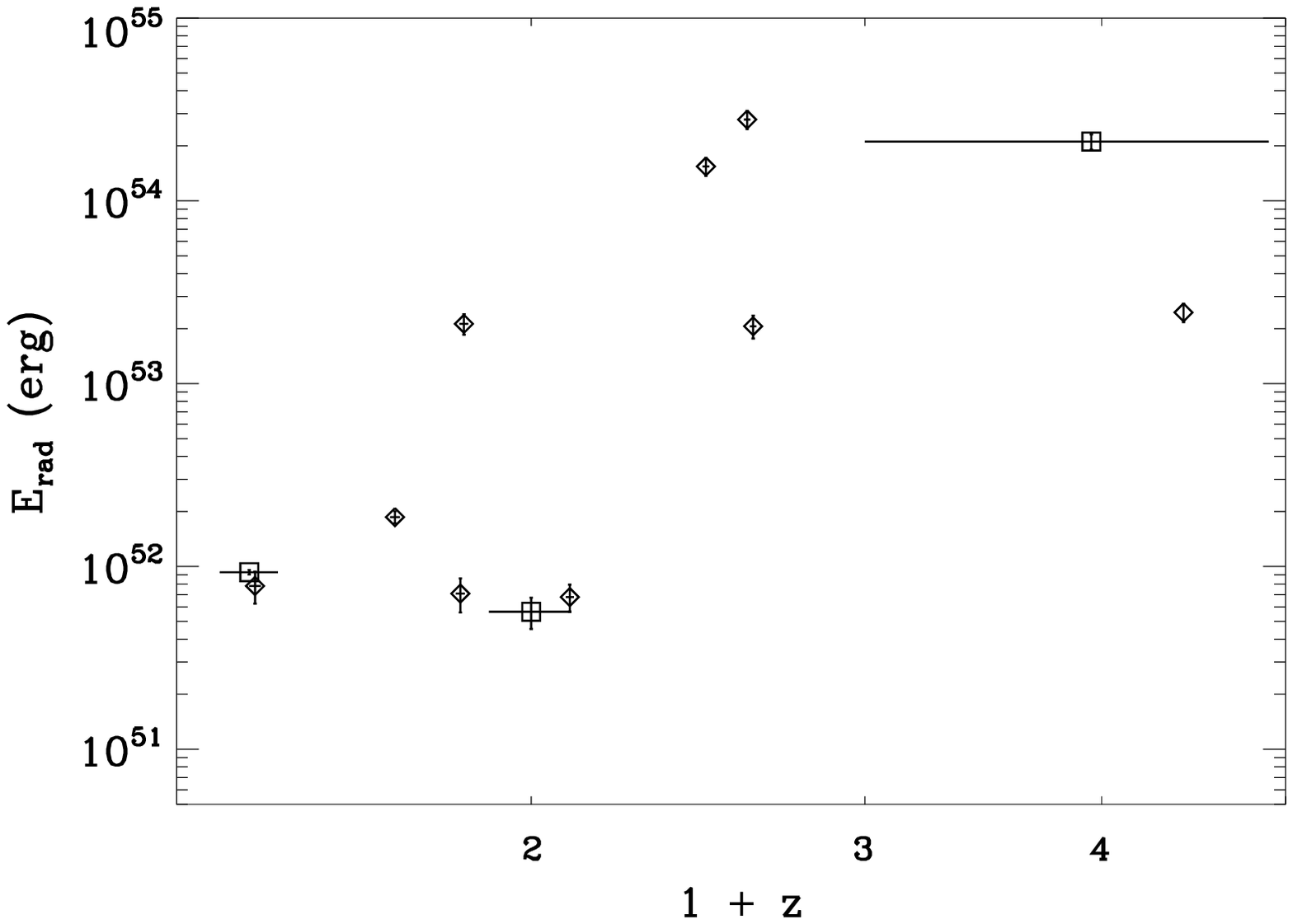}
\end{minipage}
\hfill
\caption{Left Panel: Variability Measure of Luminosity versus redshift.
\citet{lr2000}.  Right panel: $E_{\rm iso}$ versus redshift for {\it
Beppo}SAX GRB afterglows with redshift determinations.  \citet{amati2002}.
}
\label{feebleweeble}
\end{figure}

Using a Cepheid-like variability-based redshift measure developed
together with \citet{fr2000}, \citet{lr2000} found evidence for a
positive correlation between isotropic-equivalent peak luminosity $L_{\rm
iso}$ and redshift, using BATSE data.  The left panel of
Figure~\ref{feebleweeble} shows their variability-based luminosity
estimator, plotted as a function of redshift.  The diagonal lines
represent the BATSE 10\% and 90\% detection thresholds.

As can be seen in the figure, there is a dearth of high-luminosity bursts
at estimated redshifts below $z\sim 1$ and the estimated luminosities
also ``peel off'' the threshold at estimated redshifts below $z\sim 1$,
hinting at a correlation of burst luminosity with redshift. 
Nevertheless, threshold effects, together with the possibility of unknown
systematic effects intrinsic to the variability measure, are causes for
concern.

\citet{lfr2002} addressed threshold truncation effects in a study of 220
BATSE GRBs.  They found that after applying the correction, there was
significant evidence for a correlation slope $L_{\rm iso}\sim(1+z)^{1.4}$.

The number of GRBs with known redshifts has now grown to the point that
it is possible to do meaningful studies of the distributions of GRBs with
spectroscopic redshifts, as opposed to the less certain redshifts derived
from the GRB variability measure. \citet{amati2002}, using a sample of 12
{\it Beppo}SAX GRBs with afterglows and redshift estimates, found a
positive correlation between the isotropic-equivalent prompt energy
$E_{\rm iso}$ and redshift, with a reported significance of 7\%.  The
right panel of Figure~\ref{feebleweeble} shows the data used  by
\citet{amati2002}.  While the result is  promising, threshold effects
were not quantified, and the distribution of $L_{\rm iso}$ was not
investigated.

The HETE-2 bursts increase the size of the sample of GRBs with known
redshifts by a factor of 2, and therefore offer the opportunity to
confirm (or possibly refute) the Amati et al. results.  In addition, it
is interesting to extend the Amati et al. results to isotropic-equivalent 
peak luminosities ($L_{\rm iso}$), to make contact with the results of
\citet{lr2000} and of \citet{lfr2002}.

\section{The HETE-2-Augmented GRB Sample}

The HETE-2 events considered in this study are:

\begin{itemize}

\item 9 HETE-2-localized classical GRBs with afterglows and redshifts ---
GRB010921, GRB020124, GRB021004, GRB021211, GRB030226, GRB030328,
GRB030329, GRB030429, and GRB030323.

\item 2 HETE-2-localized XRFs with afterglows and redshifts --- 
XRF020903 ($z=0.25$), and XRF030723 ($z\lesssim 0.3$).

\end{itemize}

In addition to these eleven events, we consider GRB000131 ($z=4.511$), an
IPN-located, BATSE-detected GRB.  We further consider 10 events with
well-determined redshifts from the \citet{amati2002} sample.  In treating
these events, we extract values of $L_{\rm iso}$ (not otherwise reported
in \citep{amati2002}) using the values of $E_{\rm iso}$, $S$, and $F_{\rm
peak}$ provided in their paper, with due attention to cosmological
corrections.

\begin{figure}[t]
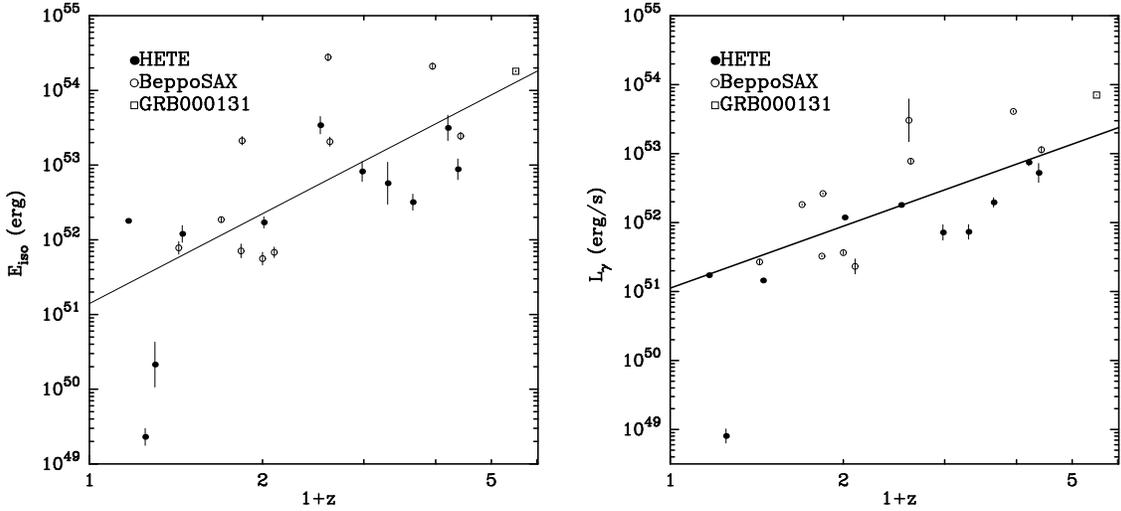


\begin{minipage}{3.0truein}
\includegraphics[width=2.8truein, clip]{Erad-z.ps}
\end{minipage}
\begin{minipage}{3.0truein}
\includegraphics[width=2.8truein, clip]{Lgamma-z.ps}
\end{minipage}
\caption{Left Panel: Isotropic-equivalent energy versus redshift.  Right
Panel: Isotropic-equivalent Luminosity versus redshift.}
\label{fooblewooble}
\end{figure}

The data are shown in Figure~\ref{fooblewooble}.  The straight lines show
fits obtained by excluding the two HETE-2-located XRFs, which would
otherwise obviously bias the fits toward very high slopes, whose value
for extrapolation to high-$z$ would be dubious.

\begin{figure}[t]
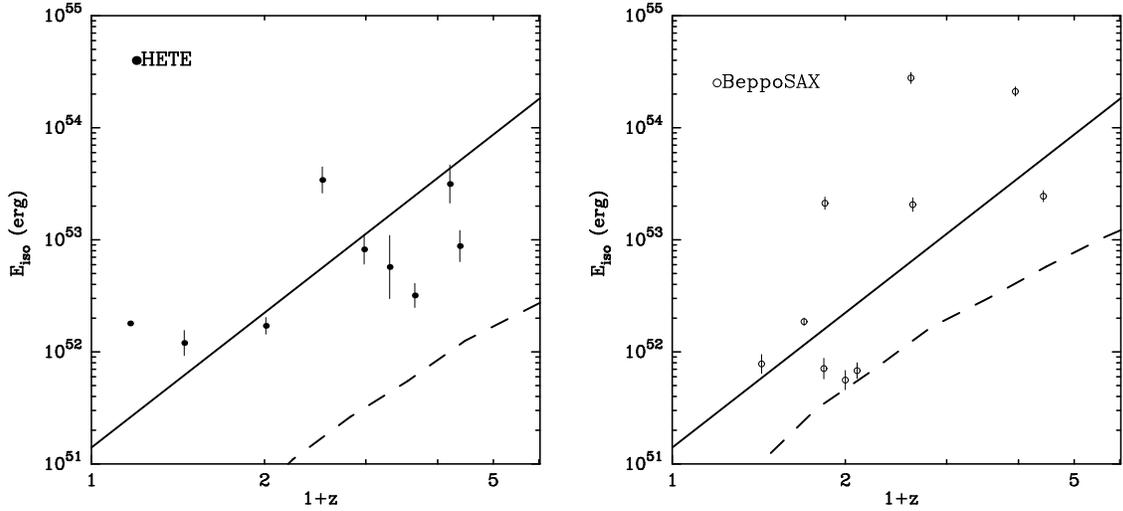


\begin{minipage}{3.0truein}
\includegraphics[width=2.8truein, clip]{eiso_1+z_hete.ps}
\end{minipage}
\begin{minipage}{3.0truein}
\includegraphics[width=2.8truein, clip]{eiso_1+z_BSAX.ps}
\end{minipage}
\caption{Isotropic-equivalent energy versus redshift.  The dashed line
shows the estimated detection threshold, calculated using the results in
\citep{band2003}.  Left Panel: HETE.  Right Panel: {\it Beppo}SAX}
\label{fibblewibble}
\end{figure}

\section{Analysis}

We perform a linear-regression analysis to the isotropic-equivalent
prompt energy and peak luminosity data, obtaining best-fit slopes and
intercepts.  We also calculate correlation coefficients, and evaluate the
formal correlation significance using the standard $t$-test.

This formal significance does not account in any way for threshold
effects. We address the question of threshold-related systematic
distortions by calculating detection thresholds using the methods
described in \citet{band2003}.  The thresholds corresponding to WXM and
to {\it Beppo}SAX are shown by the dashed lines in
Figure~\ref{fibblewibble}.

We calculate threshold-corrected correlation significance by generating
simulated samples of {\it Beppo}SAX- and HETE-2-observed data, which are
truncated by the thresholds.  The key assumption we make in the
simulations is that the burst luminosity function $f(L_{\rm iso}) \propto
L_{\rm iso}^{-1}$ \citep{dlg2003} (see also \citep{schmidt2001}).  The
significance is the fraction of the simulated samples with correlation
coefficients that exceed the observed value.

\section{Results}

The results of the analysis of the correlation of $L_{\rm iso}$ with
$1+z$ are reported in table \ref{oogabooga}, those of the analysis of the
correlation of $E_{\rm iso}$ with $1+z$ are reported in table
\ref{arglebargle}.

Threshold effects are clearly playing an important role, and reduce the
significance of the evidence for evolution substantially.  Nonetheless,
after accounting for them, there remains modest (5\% significance)
evidence for evolution of $E_{\rm iso}$, and encouraging ($9.5\times
10^{-3}$ significance) evidence for evolution of $L_{\rm iso}$.

The slopes that we find are typically 3-4.  These values are considerably
higher than that found by \citet{lfr2002}.  However, our slopes are
uncertain, and while our significances are corrected for threshold
effects, our slopes are not.  The magnitude of this effect remains to be
estimated.

\section{Discussion}

If the isotropic-equivalent energies and luminosities of GRBs really do
increase with redshift, as the evidence presented here appears to
suggest, the consequences for GRB models are profound.  Models that rely
on variations of the characteristic size of the jet opening angle to
produce the observed variations in the isotropic-equivalent energies and 
luminosities of GRBs (e.g., the uniform jet model \citep{dlg2003})
interpret the correlation as saying that higher-$z$ GRBs have narrower
jets.  This suggests that the cores of the high-$z$ progenitor stars may
rotate more rapidly than those at low $z$.  On the other hand, models
that rely exclusively on variations of viewing angle to produce the
observed variations in the isotropic-equivalent energies and luminosities
of GRBs (e.g., the universal jet model) run into difficulty explaining
why the viewing angle distribution should change as a function of
redshift.

If the isotropic-equivalent energies and luminosities of GRBs really do
increase with redshift, GRBs at very high redshifts may be more luminous
-- and therefore easier to detect -- than has been thought, which would
make GRBs a more powerful probe of cosmology and the early universe than
has been thought.

\begin{table}[t!]
\begin{tabular}{llll}
\hline
\tablehead{1}{l}{t}{Quantity} &
\tablehead{1}{c}{t}{HETE-2 + GRB000131} &
\tablehead{1}{c}{t}{{\it Beppo}SAX + GRB000131} &
\tablehead{1}{c}{t}{Combined}\\
\hline
Sample Size & 10 & 11 & 20\\
Correlation Coefficient & 0.87 & 0.83 & 0.76\\
Formal Significance & $9.0\times 10^{-4}$ & $1.7\times 10^{-3}$ &
$8.9\times 10^{-5}$\\
Threshold-Corrected &&& \\
Significance & $1.2\times 10^{-2}$ &
$4.9\times 10^{-2}$ & $9.5\times 10^{-3}$ \\
Slope$^*$ & $3.2\pm 0.5$ & $4.1\pm 0.8$ & $3.3\pm 0.6$\\
\hline
\end{tabular}
\caption{Correlation Results: $\log(L_{\rm iso})$ vs. $\log(1+z)$}
\label{oogabooga}
\end{table}
\begin{table}[t!]
\begin{tabular}{llll}
\hline
\tablehead{1}{l}{t}{Quantity} &
\tablehead{1}{c}{t}{HETE-2 + GRB000131} &
\tablehead{1}{c}{t}{{\it Beppo}SAX + GRB000131} &
\tablehead{1}{c}{t}{Combined}\\
\hline
Sample Size & 10 & 11 & 20\\
Correlation Coefficient & 0.74 & 0.75 & 0.66\\
Formal Significance & $1.3\times 10^{-2}$ & $8.1\times 10^{-3}$ &
$5.3\times 10^{-4}$\\
Threshold-Corrected &&& \\
Significance & $7.8\times 10^{-2}$ &
$1.1\times 10^{-1}$ & $5.1\times 10^{-2}$ \\
Slope$^*$ & $2.0\pm 1.0$ & $4.3\pm 1.1$ & $3.0\pm 0.8$\\
\hline
\multicolumn{4}{l}{$^*$ Uncertainties in the slope are 68\% confidence levels.}
\end{tabular}
\caption{Correlation Results: $\log(E_{\rm iso})$ vs. $\log(1+z)$}
\label{arglebargle}
\end{table}


\begin{thebibliography}{1}

\bibitem[Amati et al.(2002)]{amati2002}
Amati, L., et al. 2002, A\&A, 390, 81

%
\bibitem[Band(2003)]{band2003}
Band, D. L. 2003, ApJ, 588, 945

\bibitem[Donaghy, Lamb, \& Graziani(2003)]{dlg2003}
Donaghy, T., Lamb, D. Q., and Graziani, C. 2003, these proceedings

\bibitem[Fenimore \& Ramirez-Ruiz(2000)]{fr2000}
Fenimore, E. E., and Ramirez-Ruiz, E. 2000,  ApJ, submitted
(astro-ph/0004176)

\bibitem[Lamb \& Reichart(2000)]{lr2000}
Lamb, D.~Q. \& Reichart, D.~E. 2000, ApJ, 535, 1

\bibitem[Lloyd-Ronning, Fryer, \& Ramirez-Ruiz(2002)]{lfr2002}
Lloyd-Ronning, N. M., Fryer, C. L., and Ramirez-Ruiz, E.2000, ApJ, 574,
554

\bibitem[Schmidt(2001)]{schmidt2001}
Schmidt, M. 2001, ApJ, 552, 36

\end{thebibliography}
\end{document}